# The optimum angle-cut of collimator for the dense objects in high-energy proton radiography[*]

XU Hai-Bo（许海波）　ZHENG Na（郑娜）[1)]

Institute of Applied Physics and Computational Mathematics, Beijing 100094, China

**Abstract:** The use of the minus identity lenses with the angle-cut collimator can achieve high contrast image in high-energy proton radiography. This article presents the principle of choosing the angle-cut aperture of the collimator for the different energies and objects. The numerical simulation using the Monte Carlo code Geant4 has been implemented to investigate the entire radiography for the French test object. The optimum angle-cut apertures of the collimators are also obtained for different energies.

## 1  Introduction

High-energy proton radiography could provide a new, quantitative, and much more capable diagnostic technique to interrogate the aspects for hydrotest experiments [1-3]. The three most important effects on the protons as they go through an object are absorption, multiple Coulomb scattering (MCS), and energy loss. The MCS leading to the image blur in the radiographs, so that the proton radiography had not been used for a long time. The key technology that led to the development of proton radiography is a magnetic imaging lens system located between the object and the image which form a point-to-point focus of the proton beam and provide good position resolution over the entire field of view required for radiography, was discovered at Los Alamos National Laboratory [4]. Now, proton radiography has been shown that is far superior to flash x-ray radiography [5, 6].

The lens is an inverting, unit-magnification lens or minus identity lens. Mottershead C T and Zumbro J D demonstrated that it is possible to sort the scattered beam in terms of how it has been scattered. There is a location where the protons are sorted by the magnitude of their scattering. This always occurs in the mid-plane (Fourier plane) of a chromatically matched identity lens, where the trajectory position depends only on the MCS angle, independent of the initial position [4, 7]. If one places a angular collimator at this intermediate Fourier plane where the rays are completely sorted by MCS angle, it is possible to apply angle-cut to the proton beam, removing part of the scattered beam. After exiting this lens, the protons are focused in the spatial plane.

---

[*] Supported by NSAF (11176001) and Science and Technology Developing Foundation of China Academy of Engineering Physics (2012A0202006)
　1) E-mail：nnazheng@gmail.com.

It should also be noted that by using a single magnetic lens with just a angle-cut, one can achieve high contrast image in proton radiography. Thus by changing the aperture to provide that optimal angle-cut, one can tune the system to provide optimum sensitivity. High-energy protons of 10GeV to 50GeV can be used to radiograph dense objects. The angle-cut must be different for the different proton energies and objects. In order to obtain the best image which may produce an intensified effect between the special points or structures in transmission, it is desirable to choose the matching angular collimator and give an optimum angle-cut of collimator.

## 2  The basic principle of proton radiography

The processes for proton radiography can be descibed by assuming a simple exponential formula for nuclear attenuation and the angular distribution of scattering as Gaussian MCS [7]. In this approximation, the transmission of proton through the magnetic lens and angle-cut collimators has the following form:

$$T(L) = \exp\left(-\sum_i \frac{L_i}{\lambda_i}\right)\left[1 - \exp\left(-\frac{\theta_{\text{cut}}^2}{2\theta_0^2}\right)\right] \quad (1)$$

Here, $L_i$ is the areal density for the $i'th$ material, $\lambda_i$ is the nuclear attenuation factor for the $i'th$ material given by

$$\lambda_i = \frac{A}{N_A \sigma_i} \quad (2)$$

where $N_A$ is is Avogadro's number, $A_i$ is the atomic weight and $\sigma_i$ is the absorption cross section for the $i'th$ material.

$\theta_{\text{cut}}$ is the angle-cut imposed by the angular collimator and $\theta_0$ is the multiple coulomb scattering angle given approximately by

$$\theta_0 \approx \frac{14.1\,\text{MeV}}{pc\beta}\sqrt{\sum_i^n \frac{L_i}{X_{0i}}} \quad (3)$$

Here, $p$ is the beam momentum, $\beta = v/c$ where $v$ is the beam velocity and $c$ is the speed of light, $X_{0i}$ is the radiation length for the $i$'th material given by

$$X_{0i} = \frac{716.4 A_i}{Z_i(Z_i+1)\ln(287/\sqrt{Z_i})} \qquad (4)$$

The first term of Eq. (1) in the attenuation is the nuclear attenuation and is analogous to x-ray attenuation processes, but the second term is due to angular attenuation and makes proton radiography unique. Angular attenuation allows another way of distinguishing material properties.

For thick objects, the angular spread of the beam fraction that passes through the magnetic lens to form the image is dominated by multiple Coulomb scattering (MCS) in the radiographed object, but elastic proton–nucleon scattering in the object can also have a significant effect on the transmitted angular distribution [8, 9]. The above expressions should be correct if the angular distribution of the scattering is Gaussian MCS.

## 3 The principle of choosing the angle-cut aperture of the collimator

Different collimator apertures will permit obtaining different radiographs per proton pulse in the image plane. In order to obtain the best image which may produce an intensified effect between the special points or structures in transmission, it is desirable to choose the matching angular collimator. The multiple coulomb scattering angle $\theta_0$ is proportional to the beam energy, so that the angle-cut $\theta_{cut}$ must be different for the different proton energies. By placing an aperture restriction at the Fourier plane that have removed the scattered beam with large angles, image contrast can be enhanced to give optimal images for the object.

The transmission along ray $j$ can be expressed by

$$T(L_j) = \exp\left(-\sum_i \frac{L_{ij}}{\lambda_i}\right)\left[1 - \exp\left(-\frac{\theta_{cut}^2}{2\theta_{0j}^2}\right)\right] \qquad (5)$$

Suppose that $a$ and $b$ are the two pixels need to observant consideration in the image plane, the difference of transmission between them can be written as

$$\Delta T = \exp\left(-\sum_i \frac{L_{ia}}{\lambda_i}\right)\left[1 - \exp\left(-\frac{\theta_{cut}^2}{2\theta_{0a}^2}\right)\right] - \exp\left(-\sum_i \frac{L_{ib}}{\lambda_i}\right)\left[1 - \exp\left(-\frac{\theta_{cut}^2}{2\theta_{0b}^2}\right)\right] \qquad (6)$$

The value of the optimal cut angle can be determined by Eq. (6). At high energies where the mean free path $\lambda_i$ for the *i'th* material are approximately constant, by setting the derivative of $\Delta T$ with respect to $\theta_{cut}$ to zero, the optimum angle-cut can be obtained.

## 4  The simulation on Geant4 toolkit

### 4.1 The setup of simulation

Proton radiography requires a high energy beam to penetrate the thick objects while keeping the MCS angle and energy loss small enough to allow good spatial resolution. In this paper, we will introduce a scenario of 10 GeV, 23 GeV and 50 GeV proton radiography beamline, in which a matching section, the minus identity lens system and imaging system will be particularized [10]. The beam is first prepared with a diffuser and matching lens to meet optics requirements. Next the beam is measured just upstream of the object by the front detectors after which it passes through the object being radiographed. The beam extracted from the AGS was focused onto a 1.2 cm-thick tantalum diffuser. The lens consisted of 4 quadrupoles that were configured to form a unit magnification imaging lens. Either of a pair of collimators, 1.2 m long right circular cylinders of tungsten, was located at the Fourier mid-plane of the lens [11]. Parameters of the minus identity lens for 10 GeV, 23 GeV and 50 GeV proton radiography are given in Table 1.

Table 1.  Parameters of the minus identity lens for 10GeV, 23GeV and 50 GeV proton radiography.

| momentum /GeV·c$^{-1}$ | quadrupole aperture/mm | quadrupole gradient /T·m$^{-1}$ | quadrupole length/m | drift length/m | chromatic aberration coefficient/m | field of view/mm |
|---|---|---|---|---|---|---|
| 10 | 110 | 10 | 0.4 | 3 | 30.2 | 60 |
| 23 | 120 | 8 | 2 | 3.4 | 40.4 | 80 |
| 50 | 220 | 16.74 | 3 | 4 | 36.1 | 120 |

### 4.2 The particle tracking in the minus identity lens

The simulations of the transport of protons have been implemented with Geant4 toolkit [12] in the above proton radiography beam line. Horizontal and vertical plane ray traces of proton traveling through the minus identity lens are shown in Fig. 1. The tracing rays from object location are at angles of 0 mrad, ±2mrad, ±5mrad . The scattered protons are sorted into radial positions proportional to their polar scattering angle in the object plane. The exit beam correlations are the same as the input correlations.

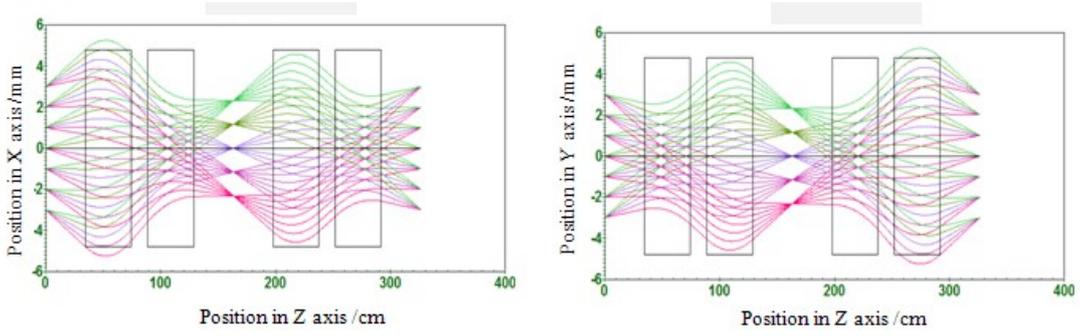

Fig. 1. Horizontal (a) and vertical (b) plane ray traces of proton traveling through the minus identity lens.

### 4.3 The optimum angle-cut of angular collimator for French test object

The central part of the collimator description in the simulation was the same as the experimental set up. The collimators approximated multiple-scattering angle acceptance cuts of 6.68 mrad [13].

We have taken data on a thick test object, the so-called French Test Object (FTO), which was designed to allow French and U.S. experimenters to collaborate on high-energy x-ray radiography methods and analysis, and their detection [14]. The FTO is a standard object used to compare radiographic capability among flash X-ray machines. The FTO consisted of three concentric spherical shells: the inner was uranium with an inside radius of 1 cm and an outside radius of 4.5 cm. This was surrounded by a copper shell of outside radius of 6.5 cm, and this was surrounded by a shell of foam plastic with an outside radius of 22.5 cm. The FTO presents a maximum areal density of 214 g/cm$^2$, where the proton ray transmits along inner edge of uranium.

Table 2. The structure of FTO and analogous FTO

| material | | air | uranium | copper | foam |
|---|---|---|---|---|---|
| radius/cm | FTO | 1.0 | 4.5 | 6.5 | 22.5 |
| | A-FTO | 2.0 | 4.5 | 6.5 | 22.5 |
| density/g·cm$^{-3}$ | | 0 | 18.9 | 8.9 | 0.5 |

Indeed, the most challenging of these hydrotest experiments use a radiographic technique to image the imploding heavy metal (uranium) with the cavity in the center. In the other words, the difference of the transmissions between central point and the points surround inner edge of uranium. The angle-cut must be different for the different proton energies and objects. In order to do comparison with different objects, an object with a small different from FTO is presented. The structures are given in Table 2.

The transmissions for FTO and analogous FTO are taken with different angle-cut apertures of the collimators. The transmission versus angle-cut aperture for 10 GeV, 23 GeV and 50 GeV proton radiography are plotted in Fig. 2. The extremum of the curve is the optimum angle-cut apertures. From Fig. 2, we can see that the optimum angle-cut apertures of the collimators are 13.09 mrad for 10 GeV, 6.68 mrad for 23 GeV and 2.56 mrad for 50 GeV for FTO, and 11.66 mrad for 10 GeV, 5.30 mrad for 23 GeV and 2.49 mrad for 50 GeV for analogous FTO.

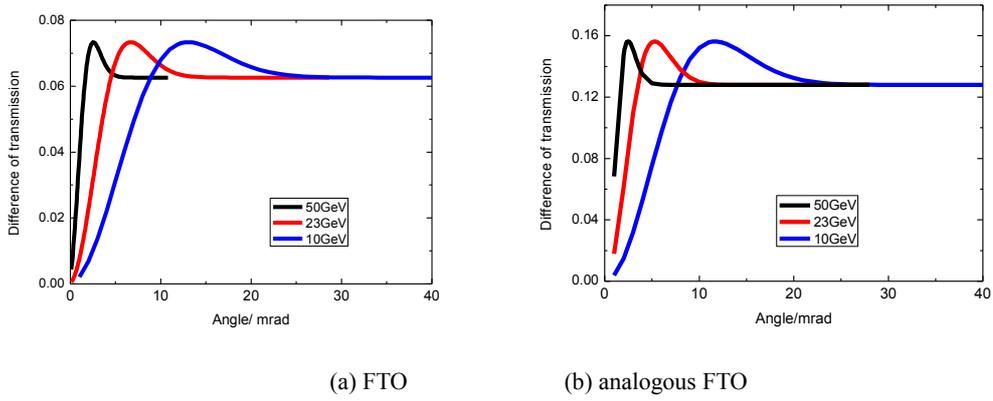

(a) FTO          (b) analogous FTO

Fig. 2. The difference of transmission between central point and the point of maximum object thickness as a function of angle-cut aperture.

### 4.4 The transmission of French test object with the optimum angle-cut aperture

The simulation results from proton radiograph image plane of FTO are shown in Figs. 3 and 4. The image in Fig. 3(a) corresponds to the optimum angle-cut apertures of the 13.09 mrad collimator, and the image in Fig. 3(b) corresponds to the normal 6.68 mrad collimator for 10 GeV, respectively. Fig. 3(c) is radial distribution for the radiograph of FTO which the most outer dosage is unitary. The image in Fig. 4(a) corresponds to the optimum angle-cut apertures of the 2.56 mrad collimator and the image in Fig. 4(b) corresponds to the normal 6.68 mrad collimator for 50 GeV respectively. Fig. 4(c) is radial distribution for the radiograph of FTO which the most outer dosage is unitary. It can been seen from Figs. 3 and 4, with the increase of angle-cut, the protons received in the image plane will be larger, and the image become lighter. For 10 GeV protons, it is obvious that all structure of the object can be seen from the image of angle-cut at 13.09 mrad but the core of the object can not be distinguished if the angle-cut at 6.68 mrad. For 50 GeV protons, the images of angle-cut at 2.56 mrad and 6.68 mrad are able to show the structure of the object, and the definition of the image of angle-cut at 2.56 mrad is a little higher than at 6.68 mrad. This conclusion is consistent with Fig. 2 (a). It is showned that, for 10 GeV protons, 6.68 mrad is in the region where the difference of transmission between central point and the point of maximum object thickness is zooming with the angle, while for 50GeV protons, 6.68 mrad is near by the peak value. Thus, the image contrast can be enhanced with the optimum angle-cut aperture collimator and a better image can be achieved with higher proton energy.

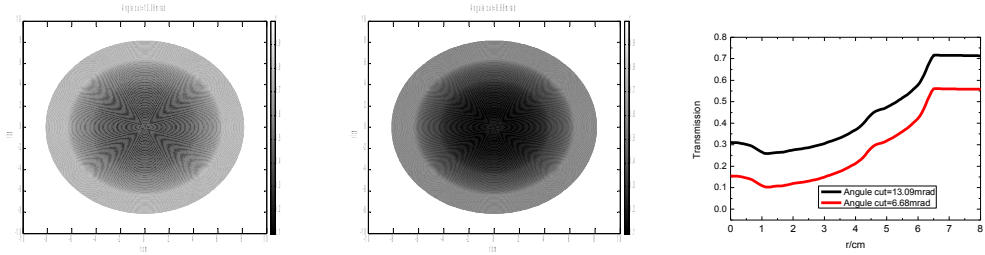

Fig. 3. Simulation results from image plane of FTO at 10 GeV protons with angle-cut (a) 13. 09 mrad, (b) 6.68 mrad, and (c) the transmission along radius.

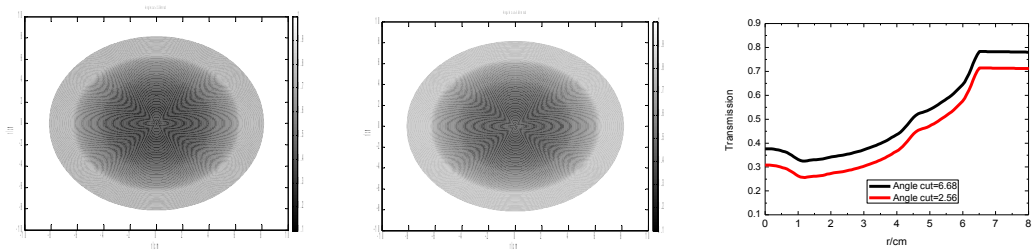

Fig. 4. Simulation results from image plane of FTO at 50 GeV protons with angle-cut (a) 2.56 mrad, (b) 6.68 mrad, ans (c) the transmission along radius.

## 5  Summary

The ability of proton radiography to adjust the image contrast by adjusting the angle-cut aperture of the collimator has been demonstrated. There is an optimum angle-cut aperture of the collimator for a given object and a given energy. The angle-cut aperture of the collimator is chosen according to the optical thickness of the object. In this article, using the Monte Carlo code Geant4, the optimum angle-cut apertures of the collimators are obtained that 13.09 mrad for 10 GeV, 6.68 mrad for 23 GeV and 2.56 mrad for 50 GeV for the French test object, respectively. The simulation results are benefited for the design of the magnetic imaging lens in high-energy proton radiography.

# 致密客体高能质子照相中准直角的优化选取

XU Hai-Bo（许海波）　　ZHENG Na（郑娜）

**摘要**　　磁透镜成像系统中角度准直器的孔径对于成像的对比度至关重要。本文根据透射率与准直角度的关系，给出了角度准直器的选择原理。并基于Geant4程序，对10 GeV、23 GeV、50 GeV高能质子照射法国试验客体进行了模拟，分别给出了不同能量下的最佳准直角度。

**关键词**：质子照相，库伦散射，角度准直器，法国试验客体，Geant4